\documentclass[10pt,twocolumn,showpacs,preprintnumbers]{revtex4}%
\usepackage{mathrsfs}
\usepackage{graphicx}
\usepackage{bm}
\usepackage{graphicx,amsmath,mathrsfs}
\usepackage{amsmath}
\usepackage{amsfonts}
\usepackage{amssymb}
\usepackage{subfigure}%
\providecommand{\U}[1]{\protect\rule{.1in}{.1in}}

\begin{document}
\title{Axially deformed relativistic Hartree Bogoliubov with separable pairing force}
\author{Yuan Tian$^{1,2,3}$, Zhong-yu Ma$^{1,2,4}$, P. Ring$^{2,3}$}
\affiliation{(1)~China Institute of Atomic Energy, Beijing 102413,
P.R.of China} \affiliation{(2)~Kavli Institute for Theoretical
Physics China, CAS, Beijing 100190, China}
\affiliation{(3)~Physikdepartment, Technische Universit\"at
M\"unchen, D-85748, Garching, Germany} \affiliation{(4)~Centre of
Theoretical Nuclear Physics, National Laboratory of Heavy Collision,
Lanzhou 730000, P.R.of China}

\begin{abstract}
A separable form of pairing interaction in the $^{1}S_{0}$ channel
has been introduced and successfully applied in the description of
both static and dynamic properties of superfluid nuclei. By
adjusting the parameters to reproduce the pairing properties of the
Gogny force in nuclear matter, this separable pairing force is
successful in depicting the pairing properties of ground states and
vibrational excitations of spherical nuclei on almost the same
footing as the original Gogny force. In this article, we extend
these investigations for Relativistic Hartree Bogoliubov theory in
deformed nuclei with axial symmetry (RHBZ) using the same separable
pairing interaction. In order to preserve translational invariance
we construct one- and two-dimensional Talmi-Moshinsky brackets for
the cylindrical harmonic oscillator basis. We show that the matrix
elements of this force can then be expanded in a series of separable
terms. The convergence of this expansion is investigated for various
deformations. We observe a relatively fast convergence. This allows
for a considerable reduction in computing time as compared to
RHBZ-calculations with the full Gogny force in the pairing channel.
As an example we solve the RHBZ equations with this separable
pairing force for the ground states of the chain of Sm-isotopes.
Good agreement with the experimental data as well as with other
theoretical results is achieved.

\end{abstract}

\maketitle

\section{Introduction}

Covariant density functional theory (DFT) based on Relativistic Mean
Field (RMF) theory provides a microscopically consistent description
of the nuclear many-body problem~\cite{LNP.641}. Conventional DFT
without particle-particle (pp) correlations can be applied only for
a few doubly closed shell nuclei. For the vast majority of nuclei
and in particular those far away from the $\beta$-stability line,
that play an important role in astrophysical applications, the
inclusion of particle-particle correlations is essential for a
quantitative description of many phenomena in nuclear structure. In
the framework of DFT pairing correlations are taken into account in
the form of Hartree-Bogoliubov
theory~\cite{RS.80,KuR.91,Ring.96,BHR.03,VALR.05} for the ground
states. Because of the numerical complexity of the deformed
Hartree-Bogoliubov equations, monopole pairing or density dependent
$\delta $-pairing interactions have been widely used in the
literature for deformed mean field calculations~\cite{BHR.03}. But
the results are affected by a cutoff parameter which has to be
introduced in a rather arbitrary way. Recent investigations show
clearly that many results depend on this cut-off~\cite{KAL.09}.
Therefore, in order to avoid the complicated problem of a pairing
cutoff, the finite range Gogny force\cite{DG.80} has been applied in
many relativistic applications~\cite{GEL.96}. The parameters of this
force have been adjusted very carefully in a semi-phenomenological
way to characteristic properties of the microscopic effective
interactions and to experimental data~\cite{BGG.84,BGG.91}. It has
been shown in several applications that the RHB model with Gogny
pairing provides an excellent tool for the description of ground
state properties in finite nuclei. Such investigations have been
first devoted to spherical applications, such as to halo phenomenon
in light nuclei~\cite{MPR.97,PVL.97}, to the properties of nuclei
near the neutron drip line~\cite{LVP.98}, to the reduction of the
spin-orbit potential in nuclei with extreme isospin
values~\cite{LVP.98a}, to ground-state properties of Ni and Sn
isotopes~\cite{LVR.98}, etc. RHB theory with Gogny pairing has also
been applied to investigate
deformed~\cite{LVRS.99,LVR.99,LVR.99a,AKR.99} and
rotating~\cite{AKR.99} nuclei, but such calculations are limited due
to their numerical complicity.

Recently we have introduced a new separable form of the pairing
force for RHB calculations in spherical nuclei~\cite{TMR.09a,TM.06}.
The parameters of this separable force are adjusted to reproduce the
pairing properties of the Gogny force in nuclear matter. It
preserves translational invariance and has finite range. A similar
ansatz has been used in the pairing channel of non-relativistic
Skyrme calculations in Refs.~\cite{DL.08,LDB.09}. This pairing
interaction is separable in momentum space. In $r$-space the
translational invariance leads to a $\delta$-force in the center of
mass coordinates and therefore, at a first glance, translational
invariance forbids exact separability. However using well known
techniques of Talmi and Moshinsky~\cite{Tal.52,Mos.59,BJM.60} it has
been shown in Ref.~\cite{TMR.09a} that this force can be represented
by a sum of separable terms which converges quickly. This avoids the
complicated problem of a cutoff at large momenta or energies
inherent in zero range pairing forces. As we discussed in Fig.~6 of
Ref.~\cite{TMR.09a}, we found that although the $\delta$-force can
give the same average gap as the Gogny force D1S if the size of the
strength is adjusted properly, the individual matrix elements of the
forces and the matrix elements of the pairing field $\Delta$ are
very different from each other. The $\delta$-force behaves in some
sense very much like a constant pairing. In contract our separable
force has a very similar behavior to the Gogny force.

This simple separable force can reproduce the pairing properties of
the ground-state for spherical nuclei on almost the same footing as
the original Gogny pairing interaction. Recently it has also been
applied for studying dynamic properties of spherical
nuclei~\cite{TMR.09b}. The relativistic quasiparticle random phase
approximation (RQRPA) based on the same separable pairing
interaction in the pairing channel was used for calculations of
low-lying 2$^{+}$ and 3$^{-}$ excited states in a chain of
Sn-isotopes, which are very sensitive to the pairing channel. In
comparison with experimental data and with the results of the
original Gogny force it was shown that this simple separable pairing
interaction is also very successful in depicting the dynamical
pairing properties of vibrational excitations.

So far, the new separable pairing force has been used only in
spherical nuclei. In this article we apply this separable pairing
force for axially deformed relativistic Hartree-Bogoliubov (RHBZ)
calculations. For axially symmetric shapes the densities are
invariant with respect to a rotation around the symmetry axis, which
is taken to be the z-axis. Therefore it is convenient to work in
cylindrical coordinates. The Talmi and Moshinsky
techniques~\cite{Tal.52,Mos.59,BJM.60} used in Ref.~\cite{TMR.09a}
are restricted to spherical coordinates. Therefore we had to develop
similar techniques for cylindrical coordinates working in an
anisotropic oscillator basis. Again the matrix elements of this
pairing force in this basis are no longer fully separable. However
they can be expanded, as in the spherical case, in a series of
separable terms. Obviously the convergence of this expansion is not
as fast as in the spherical case, it is still quick enough to save
considerable numerical effort as compared to the full Gogny
calculations. Finally we investigate for a chain of Sm-isotopes the
ground-state properties such as binding energies and deformations
using the RHBZ-program with this new separable pairing force. As it
is known, the ground-state properties of the deformed nuclei are
highly affected by the pairing gap. Good agreement is found, when
comparing with experimental data and with theoretical calculations
using RMF + BCS theory with a constant pairing force, or using
non-relativistic Hartree-Fock Bogoliubov (HFB) theory with the full
Gogny force D1S.

The paper is arranged as follows. The theoretical formalism of RHBZ
with the separable form of the pairing interaction is presented in
Sec.~II. The convergence of the expansion of the pairing force
elements in the cylindrical harmonic oscillation basis is
investigated in Sec.~III. In Sec.~IV the ground-state properties of
a chain of Sm-isotopes are calculated in the RHBZ approach. They are
discussed in Sec.~IV. Finally we give a brief summary in Sec.~V.

\section{Theoretical Formalism}

We start our investigations in symmetric nuclear matter with various
densities. The gap equation in the $^{1}S_{0}$ channel has the form,
\begin{equation}
\Delta(k)=-\int_{0}^{\infty}\frac{k^{\prime2}dk^{\prime}}{2\pi^{2}}\langle
{k}|V^{^{1}S_{0}}|k^{\prime}\rangle\frac{\Delta(k^{\prime})}{2E(k^{\prime})}~,
\label{gapeq}%
\end{equation}
where a separable form of the pairing force is
introduced~\cite{TMR.09a},
\begin{equation}
\langle{k}|V_{\mathrm{sep}}^{^{1}S_{0}}|k^{\prime}\rangle=-Gp(k)p(k^{\prime
})~. \label{sp}%
\end{equation}
A simple Gaussian ansatz $p(k)=e^{-a^{2}k^{2}}$ is assumed. In
Ref.~\cite{TMR.09a} the two parameters $G$ and $a$ have been fitted
to the density dependence of the gap at the Fermi surface
$\Delta(k_{F})$. Comparing with the Gogny force, we found two sets
of parameters $G=738$ MeV$\cdot $fm$^{3}$ and $a=0.636$ fm for the
parameter set D1~\cite{DG.80} and $G=728$ MeV$\cdot$fm$^{3}$ and
$a=0.644$ fm for the set D1S~\cite{BGG.91}.

In the Hartree approximation for a consistent mean field, the RHB
equations~\cite{KuR.91,Ring.96} read
\begin{equation}
\label{rhb}\left(
\begin{array}
[c]{cc}%
\hat{h}_{D}-\lambda & \hat{\Delta}\\
-\hat{\Delta}^{*} & -\hat{h}_{D}+\lambda
\end{array}
\right)  \left(
\begin{array}
[c]{c}%
U_{k}({\mathbf{r}})\\
V_{k}({\mathbf{r}})
\end{array}
\right)  =E_{k} \left(
\begin{array}
[c]{c}%
U_{k}({\mathbf{r}})\\
V_{k}({\mathbf{r}})
\end{array}
\right)  ~,
\end{equation}
where $\hat{h}_{D}$ is the single nucleon Dirac Hamiltonian,
\begin{align}
\label{Dirac}\hat{h}_{D}=  &  -i\mbox{\boldmath$\alpha$}\cdot
{\boldmath{\ \nabla}}+\beta(m+g_{\sigma}\sigma({\mathbf{r}}))+g_{\omega}%
\tau_{3}\omega^{0}({\mathbf{r}}) +g_{\rho}\rho^{0}({\mathbf{r}})\nonumber\\
&  +e\frac{(1-\tau_{3})}{2}A^{0}({\mathbf{r}})-m~.
\end{align}
The Dirac Hamiltonian contains the mean-field potentials of the
isoscalar scalar $\sigma$-meson, the isoscalar vector
$\omega$-meson, the isovector vector $\rho$-meson, as well as the
photon. $m$ is the nucleon mass and the term $-m$ subtracts the
rest-mass and normalizes the energy scale to the continuum limit.
The chemical potential is to be determined by the subsidiary
particle number condition, where the expectation value of the
particle number operator in the ground state equals the number of
nucleons. The column vectors are the quasiparticle spinors and
$E_{k}$ are the quasiparticle energies. $\hat{\Delta}$ is the
pairing fields, which is an integral operator with the kernel
\begin{equation}
\label{pairing}\Delta_{ab}({\mathbf{r}},{\mathbf{r}}^{\prime})=\frac{1}{2}%
\sum_{c,d}V_{abcd}({\mathbf{r}},{\mathbf{r}}^{\prime})\kappa_{cd}({\mathbf{r}%
},{\mathbf{r}}^{\prime})~,
\end{equation}
where $a,b,c,d$ denote the quantum numbers that specify the Dirac
indices of the spinor. They run over the two spin orientations and
the large and small components.
$V_{abcd}({\mathbf{r}},{\mathbf{r}}^{\prime})$ are matrix elements
of two-body pairing interaction. In general this should be a
relativistic force~\cite{KuR.91} and involve large and small
components. However, since pairing correlations in nuclei are a
purely non-relativistic effect it has been shown in
Ref.~\cite{SR.02} that we can neglect the pairing matrix elements
between large and small components as well as the effect of the
pairing matrix elements between small components and we consider
only the upper part of the pairing field $\Delta$ in Eq.
(\ref{rhb}). The pairing tensor is defined as
\begin{equation}
\label{tensor}\kappa_{cd}({\mathbf{r}},{\mathbf{r}}^{\prime})=\sum_{E_{k}%
>0}U_{ck}({\mathbf{r}})^{*}V_{dk}({\mathbf{r}}^{\prime})~.
\end{equation}

For the axially symmetric deformed shape rotational symmetry is
broken and therefore the total angular momentum $J$ is no longer a
good quantum number. However, the densities are still invariant with
respect to a rotation around the symmetry axis, which is taken to be
the z-axis. It then turns out to be useful to work with the
cylindrical coordinates
\begin{equation}
\label{E7}\pmb{r}=(r_{\bot}\cos\varphi,r_{\bot}\sin\varphi,z)~.
\end{equation}
In these coordinates the Dirac equation can be reduced to a coupled
set of partial differential equations in the two variables $z$ and
$r_{\bot}$ that are solved by an expansion in an anisotropic
harmonic oscillator basis~\cite{GRT.90}.

Since the interaction in the particle-hole (ph)-channel is identical
to earlier calculations, here we only discuss the derivation of the
matrix elements of the pairing interaction for the separable form of
Eq.~(\ref{sp}) in the pp-channel. First, we transform the separable
force in Eq.~(\ref{sp}) from momentum space to coordinate space and
obtain
\begin{equation}%
\begin{split}
V(\pmb{r}_{1},  &  \pmb{r}_{2},\pmb{r}_{3},\pmb{r}_{4})~=~-G\delta
(\pmb{R}-\pmb{R}^{\prime}) \frac{1}{2}(1-P_{\sigma})P(r)P(r^{\prime})\\
\end{split}
~ \label{vr}%
\end{equation}
where ${{\pmb{R}}}=\frac{1}{2}({{\pmb{r}}}_{1}+{{\pmb{r}}}_{2})$ and
${{\pmb{r}}}={{\pmb{r}}}_{1}-{{\pmb{r}}}_{2}$ are the center of mass
and relative coordinates of two paired particles, respectively. And
$P(r)$ is obtained from the Fourier transform of $p(k)$,
\begin{equation}
P(r)=P(z,r_{\bot})= \frac{1}{(4\pi a^{2})^{3/2}}e^{-\frac{z^{2}+r_{\bot}^{2}%
}{4a^{2}}}~. \label{gauss-r}%
\end{equation}
The term $\delta({\pmb{R}}-{\pmb{R}}^{\prime})$ in Eq.~(\ref{vr})
insures the translational invariance. It also shows that this force
is not completely separable in coordinate space.

We start from the eigenfunctions of the deformed harmonic
oscillator:
\begin{equation}%
\begin{split}
|\alpha\rangle &  =|n_{z},n_{r},m_{l},m_{s}\rangle\\
&
=\frac{1}{\sqrt{b_{z}b_{\bot}^{2}}}\phi_{n_{z}}(\frac{z}{b_{z}})\phi
_{n_{r}}^{m_{l}}(\frac{r_{\bot}}{b_{\bot}})\frac{1}{\sqrt{2\pi}}%
e^{im_{l}\varphi}\chi_{m_{s}}(s)~,
\end{split}
\end{equation}
with
\begin{equation}%
\begin{split}
\phi_{n_{z}}(x)  &  =\mathcal{N}_{n_{z}}H_{n_{z}}(x)e^{-\frac{x^{2}}{2}}~,\\
\phi_{n_{r}}^{m_{l}}(x)  &  =\mathcal{N}_{n_{r}m_{l}}\sqrt{2}x^{|m_{l}%
|}L_{n_{r}}^{|m_{l}|}(x^{2})e^{-\frac{x^{2}}{2}}~.
\end{split}
\end{equation}
The quantity $b_{z}=\sqrt{\hbar/m\omega_{z}}$ and $b_{\bot}=\sqrt
{\hbar/m\omega_{\bot}}$ are the harmonic oscillator length. The
polynomials $H_{n}(\zeta)$ and $L_{n}^{m}(\eta)$ are Hermite
polynomials and associated Laguerre polynomials as defined in
Ref.~\cite{AS.70}. The normalization constants are given by
\begin{equation}
\mathcal{N}_{n_{z}}=\frac{1}{\sqrt{\sqrt{\pi}2^{n_{z}}n_{z}!}}~,\hspace
{0.5cm}\mathcal{N}_{n_{r}m_{l}}=\sqrt{\frac{n_{r}!}{(n_{r}+|m_{l}|)!}}~,
\end{equation}
where the $m_{l}$ and $m_{s}$ are the components of the orbital
angular momentum and of the spin along the symmetry axis. The
eigenvalue of $j_{z}$, which is a conserved quantity in these
calculations, is $\Omega=m_{l}+m_{s}$, the parity is given by
$\pi=(-)^{n_{z}+m_{l}}$. The conventional deformation parameter
$\beta$ is obtained from the calculated quadrupole moments through
\begin{equation}
Q=Q_{n}+Q_{p}=\sqrt{\frac{16\pi}{5}}\frac{3}{4\pi}AR_{0}^{2}\beta~,
\label{E13}%
\end{equation}
with $R_{0}=1.2A^{1/3}$~fm. Where The quadrupole $Q_{n,p}$ moments
for neutrons and protons are calculated using the expressions
\begin{equation}
Q_{n,p}=\langle2r^{2}P_{2}(\cos\theta)\rangle_{n,p}=\langle2z^{2}-x^{2}%
-y^{2}\rangle_{n,p}~.
\end{equation}
In the pairing channel, although the total angular momentum $J$ is
no longer a good quantum number, we still have
$\Omega_{1}+\Omega_{2}=0$. Together with the projector
$\frac{1}{2}(1-P_{\sigma})$, the two-particle wave function can be
written as:
\begin{equation}
|\alpha_{1},\alpha_{2}\rangle=|n_{z_{1}},n_{z_{2}}\rangle|n_{r_{1}}m_{l_{1}%
}m_{s_{1}},n_{r_{2}}-m_{l_{1}}-m_{s_{1}}\rangle~.
\end{equation}
As it is known, the separable force in Eq.(\ref{vr}) is expressed in
the center of mass frame, the two-particle wave function has to be
transformed into the same frame. So we use the fact that the product
of two oscillator functions in the coordinates of the two particles
can be expanded in terms of products of oscillator functions in the
relative and in the center of mass coordinates. This fact holds for
the one-dimensional oscillators in $z$-direction
\[
|n_{z_{1}}n_{z_{2}}\rangle=\sum_{N_{z}n_{z}}M_{N_{z}n_{z}}^{n_{z_{1}}n_{z_{2}%
}}|N_{z}n_{z}\rangle~,
\]
as well as for the two-dimensional oscillators in
$r_{\perp}$-direction
\[
|n_{r_{1}}m_{l_{1}}n_{r_{2}}m_{l_{2}}\rangle=\sum_{N_{p}M_{p}}\sum_{n_{p}%
m_{p}}M_{N_{p}M_{p}n_{p}m_{p}}^{n_{r_{1}}m_{l_{1}}n_{r_{2}}m_{l_{2}}}%
|N_{p}M_{p}n_{p}m_{p}\rangle~.
\]
Using the multinomial coefficients defined as
\begin{equation}
\left(
\begin{array}
[c]{cccc}%
\multicolumn{4}{c}{n}\\
m_{1} & m_{2} & \dots & m_{\nu}%
\end{array}
\right)  =\frac{n!}{m_{1}!m_{2}!\dots m_{\nu}!}~, \label{mutinomial}%
\end{equation}
with $n=m_{1}+m_{2}+\dots+m_{\nu}$ for $\nu=2$ (binomial
coefficients, which is expressed as $\left(
\begin{array}
[c]{c}%
n\\
m_{1}%
\end{array}
\right)  $ and $\nu=4$, the one- and two-dimension Talmi-Moshinski
transform brackets can be written as~\cite{CL.04}: \begin{widetext}
\begin{equation}\label{}
\begin{split}
M^{n_{z_1}n_{z_2}}_{N_z~n_z}&=\frac{1}{\sqrt{2^{N_z+n_z}}}\sqrt{\frac{n_{z_1}!n_{z_2}!}{N_z!n_z!}}
\delta_{n_{z_1}+n_{z_2},N_z+n_z}
\sum_{s=0}^{n_z}(-)^s\left(\begin{array}{c}N_z\\
n_{z_1}-n_z+s\end{array}\right) \left(\begin{array}{c}n_z\\
s\end{array}\right)
\end{split}~,
\end{equation}
\begin{equation}\label{}
\begin{split}
M^{n_{r_1}m_{l_1}n_{r_2}m_{l_2}}_{N_p~M_p~n_p~m_p}%
&=\frac{(-)^{N_p+n_p-n_{r_1}-n_{r_2}}}{\sqrt{2^{2N_p
+2n_p+|M_p|+|m_p|}}}\sqrt{\frac{(n_{r_1})!(n_{r_1}+|m_{l_1}|)!(n_{r_2})!(n_{r_2}+|m_{l_2}|)!}
{(N_p)!(N_p+|M_p|)!(n_p)!(n_p+|m_p|)!}}\\
&\times\delta_{2n_{r_1}+|m_{l_1}|+2n_{r_2}+|m_{l_2}|,2N_p+|M_p|+2n_p+|m_p|}
\delta_{m_{l_1}+m_{l_2},M_p+m_p}\\
\times\sum_{Q,R,S=0}^{N_p}\sum_{T=0}^{M_p}
\sum_{q,r,s=0}^{n_p}\sum_{t=0}^{m_p}
&
(-)^{r+s+t}\left(\begin{array}{cccc}\multicolumn{4}{c}{N_p}\\
N_p-Q-R-S&~Q&~R&~S\end{array}\right)
\left(\begin{array}{c}M_p\\ T\end{array}\right)\left(\begin{array}{cccc}\multicolumn{4}{c}{n_p}\\
n_p-q-r-s&~q&~r&~s\end{array}\right) \left(\begin{array}{c}m_p\\
t\end{array}\right)~.
\end{split}
\end{equation}
Therefore the matrix element of the separable force in axially
deformed oscillator basis can be written as:
\begin{equation}
\langle12|V|1^{\prime}2^{\prime}\rangle=\langle n_{z_{1}}n_{r_{1}}m_{l_{1}%
},n_{z_{2}}n_{r_{2}}m_{l_{2}}|V|n_{z_{1^{\prime}}}n_{r_{1^{\prime}}%
}m_{l_{1^{\prime}}},n_{z_{2^{\prime}}}n_{r_{2^{\prime}}}m_{l_{2^{\prime}}%
}\rangle\label{vs}\\
=-G\,\sum_{N_{z}}^{N_{z}^{0}}\sum_{N_{p}}^{N_{p}^{0}}\,W_{1^{{}}2}^{N_{z}%
N_{p}}\cdot W_{1^{\prime}2^{\prime}}^{N_{z}N_{p}}~,
\end{equation}
\end{widetext}
where
\begin{equation}
W_{12}^{N_{z},N_{p}}=\frac{1}{b_{z}^{1/2}b_{\bot}}\frac{1}{8\pi\sqrt[4]{2}%
}1V_{1^{{}}2}^{N_{z}}V_{1^{{}}2}^{N_{p}},
\end{equation}
with
\begin{equation}%
\begin{split}
V_{12}^{N_{z}}  &  =M_{N_{z}n_{z}}^{n_{z_{1}}n_{z_{2}}}\frac{1}{\alpha_{z}%
^{{}}}%
{\displaystyle\int\limits_{-\infty}^{+\infty}}
\phi_{n_{z}}(x)e^{-\frac{x^{2}}{4\alpha_{z}^{2}}}dx~,\\
V_{12}^{N_{p}}  &  =M_{N_{p}~~0~~n_{p}~~0}^{n_{r_{1}}m_{l_{1}}n_{r_{2}%
}m_{l_{2}}}\frac{1}{\alpha_{p}^{2}}%
{\displaystyle\int\limits_{0}^{\infty}}
\phi_{n_{p}}^{0}(x)e^{-\frac{x^{2}}{4\alpha_{p}^{2}}}xdx,
\end{split}
\label{E19}%
\end{equation}
with the definition of
\begin{equation}%
\begin{split}
n_{z}  &  =n_{z_{1}}+n_{z_{2}}-N_{z}~,\\
n_{p}  &  =n_{r_{1}}+n_{r_{2}}+|m_{l_{1}}|-N_{p}~.
\end{split}
\end{equation}
and $\alpha_{z}=a/b_{z}$, $\alpha_{p}=a/b_{\bot}$. Thus we find that
the pairing matrix elements for the separable pairing interactions
used in the RHB equation can be evaluated by a sum of separable
terms in Eq. (\ref{vs}). Using this expressions the pairing field
$\Delta$ in Eq. (\ref{pairing}) has the form
\begin{equation}
\Delta_{12}=-G\,\sum_{N_{z}}^{N_{z}^{0}}\sum_{N_{p}}^{N_{p}^{0}}\,W_{1^{{}}%
2}^{N_{z}N_{p}^{\ast}}P_{N_{z}N_{p}} \label{E10}%
\end{equation}
with
\begin{equation}
P_{N_{z}N_{p}}=\frac{1}{2}\sum_{12}\,W_{1^{{}}2}^{N_{z}N_{p}}\kappa_{12}^{{}}
\label{E11}%
\end{equation}
The results of the RHBZ model will depend on the choice of the
effective RMF Lagrangian in the ph-channel, as well as on the
treatment of pairing correlations. In this work the effective
interaction NL3~\cite{LKR.97} is adopted for the RMF Lagrangian and
in the pairing channel we use the separable form of the pairing
force in Eq.~(\ref{vr}) adjusted to the pairing part of the Gogny
D1S force in Ref.~\cite{TMR.09a}.

\section{study of convergence}

In the following investigations we solve the RHBZ equation
(Eq.~\ref{rhb}) with this separable pairing force. As in
Ref.~\cite{RGL.97} the Dirac spinors are expanded in an axially
deformed oscillator basis with $N_{F}=20$ major oscillator shells.
As we see from the Eq.~(\ref{vs}) the separable pairing interaction
is not fully separable in the axially deformed harmonic oscillator
basis. We have a sum over the quantum number $N_{z}$ and $N_{p}$
characterizing the major shells of the deformed harmonic oscillator
in the center of mass coordinate. For the self-consistent
calculation of the RHBZ equation, the matrix elements of the
$V_{N_{z}}^{12}$ and $V_{N_{p}}^{12}$ are calculated and stored in
memory before the iteration. Therefore the time spend on the later
calculations of the pairing matrix elements is really negligible as
compared to the total time, while it takes a large percentage of
computing time for the RHBZ model with the fully Gogny force in the
pairing channel. Due to this big advantage much computer time can be
saved in RHBZ calculations for axially deformed nuclei.

\begin{figure}[ptb]
\includegraphics[height=2.5in]{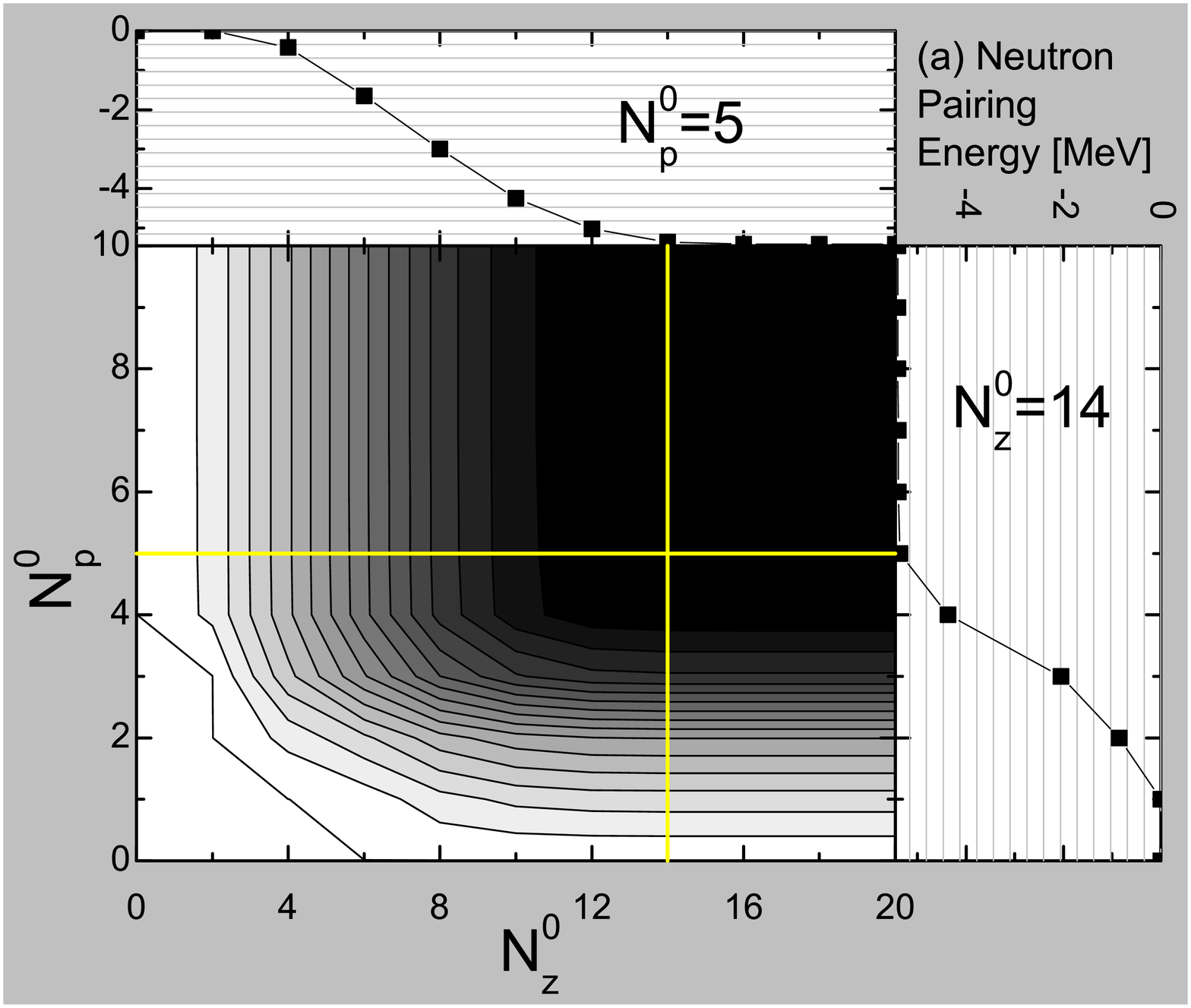}\quad\quad
\includegraphics[height=2.5in]{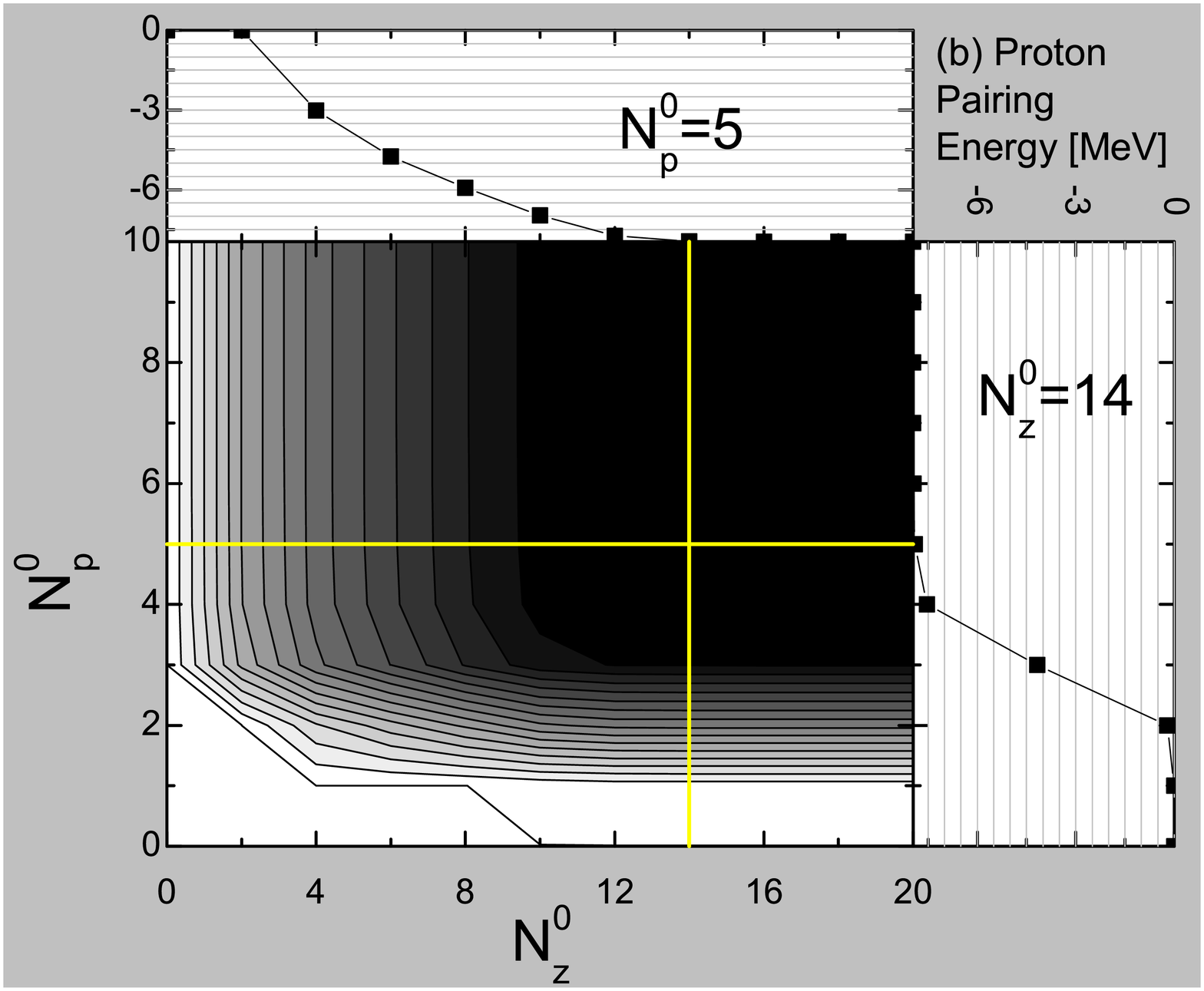}\caption{(Color online) Neutron and
Proton pairing energies of $^{164}$Er obtained with different
numbers of
separable terms $N_{z}^{0}$ and $N_{p}^{0}$ in Eq.~(\ref{vs}).}%
\label{fig1}%
\end{figure}

For $N_{F}=20$ the maximum number of $N_{z}$ and $N_{p}$ for the
expansion of
the pairing matrix elements in Eq.~(\ref{vs}) is $N_{z}^{0}=40$ and $N_{p}%
^{0}=20$ respectively, which means a large number of 8000 separable
terms in the pairing channel. As we mentioned in last section,
therefore a large memory space is required for this series. To
reduce the storage we study the convergence of the expansion with
the number of separable terms $N_{z}^{0}$ and $N_{p}^{0}$ and take
the nucleus $^{164}$Er as example. For the self-consistent solution
of the RHBZ equations with the full expansion of the separable
pairing matrix elements the deformation of the ground state of
$^{164}$Er is found to be prolate with $\beta=0.325$. Varying the
numbers $N_{z}^{0}$ and $N_{p}^{0}$ of the expansion we plot in
Fig.~\ref{fig1} the neutron and proton pairing energies of
$^{164}$Er with respect to the values of $N_{z}^{0}$ and
$N_{p}^{0}$. In the main panel of Figs.~\ref{fig1}(a) and
~\ref{fig1}(b), the contour lines correspond to lines of constant
neutron and proton pairing energy. The value of these quantities
increases with darkness.
It is clearly seen that the neutron pairing energy converges for $N_{z}%
^{0}=14$ and $N_{p}^{0}=5$. The top and right panels of
Fig.~\ref{fig1}(a) show how the neutron pairing energy converges as
a function of $N_{z}^{0}$ and $N_{p}^{0}$ for fixed values of
$N_{p}^{0}=5$ and $N_{z}^{0}=14$, respectively. The same is shown
in Fig.~\ref{fig1}(b) for the proton pairing energy. Due to the
prolate shape
of $^{164}$Er the convergence in $N_{z}^{0}$ is slower than that in $N_{p}%
^{0}$.

\begin{figure}[ptb]
\includegraphics[width=3.2in]{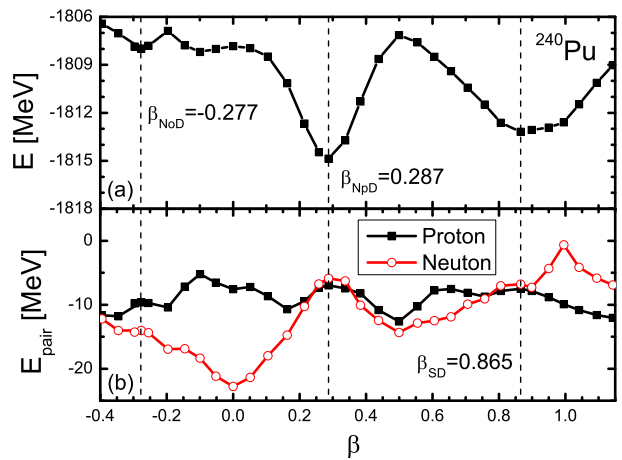}\caption{(Color online) The binding
energy (a) and the pairing energy (b) as a function of the
deformation for the nucleus $^{240}$Pu obtained with different
numbers of separable terms
$N_{z}^{0}$ and $N_{p}^{0}$ in Eq.~(\ref{vs}).}%
\label{fig2}%
\end{figure}

in Fig.~\ref{fig2}(a) the potential energy surface (PES) of the
heavy nucleus $^{240}$Pu is plotted as a function of the deformation
parameter $\beta$. The
full pairing matrix elements Eq.~(\ref{vs}) with $N_{z}^{0}=40$ and $N_{p}%
^{0}=20$ are adopted in these calculations. The corresponding
pairing energies of protons and neutrons are given in
Fig.~\ref{fig2}(b). It is found that the ground state of $^{240}$Pu
has a normal prolate deformation (NpD) at
$\beta_{\mathrm{NpD}}=0.278$. Further two minima in the PES are
observed with a normal oblate deformation (NoD) at
$\beta_{\mathrm{NoD}}=-0.277$, and a super prolate deformation (SD)
at $\beta_{\mathrm{SD}}=0.865 $. As discussed
above, a reduced number of separable terms with $N_{p}^{0}=5$ and $N_{z}%
^{0}=14$ in Eq.~(\ref{vs}) is large enough to obtain convergence for
the pairing matrix elements at the normal prolate deformation. We
also investigate the convergence with the number $N_{z}^{0}$ and
$N_{p}^{0}$ of separable terms for the other two minima in the PES.
We find that for the super deformation a larger value of
$N_{z}^{0}=18$ in the direction of the symmetry axis is needed to
obtain convergence for the global properties while in the
perpendicular direction $N_{p}^{0}=5$ is already large enough. On
the other side, for the normal oblate deformation, $N_{z}^{0}=10$ is
large enough, and we need at least $N_{p}^{0}=7$ to reach the full
convergence. The details of these results are given in the
Tab.~\ref{tab1}.

\begin{table}[ptb]
\tabcolsep 6pt
\par
\begin{center}%
\begin{tabular}
[c]{c|c|c|c}\hline $\beta_{ND}=-0.277$ & E [MeV] & $E_{n}$ [MeV] &
$E_{p}$ [MeV]\\\hline
full & -1807.984 & -14.025 & -9.639\\
$N_{z}=10$, $N_{p}=5$ & -1807.123 & -8.870 & -8.718\\
$N_{z}=10$, $N_{p}=6$ & -1807.819 & -13.022 & -9.609\\
$N_{z}=10$, $N_{p}=7$ & -1807.972 & -13.975 & -9.606\\\hline\hline
$\beta_{SD}=0.865$ & E [MeV] & $E_{n}$ [MeV] & $E_{p}$ [MeV]\\\hline
full & -1813.181 & -6.787 & -7.515\\
$N_{z}=14$, $N_{p}=5$ & -1812.896 & -5.132 & -6.462\\
$N_{z}=16$, $N_{p}=5$ & -1813.066 & -5.906 & -7.163\\
$N_{z}=18$, $N_{p}=5$ & -1813.148 & -6.452 & -7.460\\\hline
\end{tabular}
\end{center}
\caption{The the total energy $E$ and the pairing energies $E_{n}$
and $E_{p}$ for protons and neutrons for the nucleus $^{240}$Pu at
normal and at super
deformation for various values of $N_{z}^{0}$ and $N_{p}^{0}$.}%
\label{tab1}%
\end{table}

In practical applications for normal deformed prolate nuclei it
turns out that the expansion of the pairing matrix elements in
Eq.~(\ref{vs}) can be
restricted to finite values $N_{z}\leq N_{z}^{0}=14$ and $N_{p}\leq N_{p}%
^{0}=5$ for obtain sufficient accuracy. For specific cases, where
higher precision is required or for very large deformations, e.g.
for superdeformed configurations larger values for $N_{z}^{0}$ or
$N_{p}^{0}$ are required and convergence has to be checked. The
matrix elements $V_{12}^{N_{z}}$ and $V_{12}^{N_{z}}$ are calculated
and stored before starting the iterations. As compared to the
calculations with the full expansion of the separable force in the
pairing channel this corresponds to a considerable reduction in the
memory and computing time.

\bigskip

\section{Sm-isotopes}

\begin{figure}[ptb]
\includegraphics[width=3.2in]{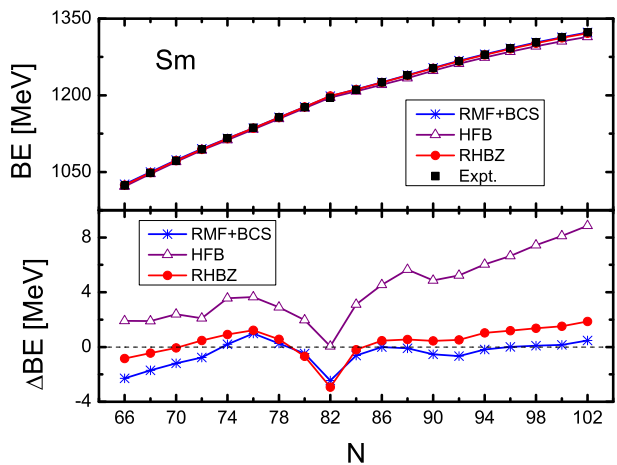}\caption{(Color online) Dependence of
the binding energy (BE) and discrepancy between calculated binding
energies and the available experimental data $\Delta$BE on the
number of neutrons for a
chain of Sm(Z=62) isotopes.}%
\label{fig3}%
\end{figure}

\begin{figure}[b]
\includegraphics[width=3.4in]{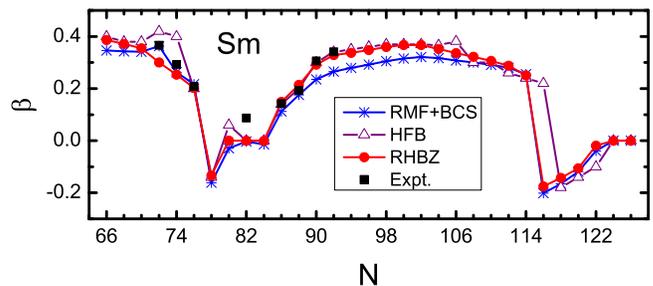}\caption{(Color online) Dependence of
the deformation $\beta$ on the number of neutrons for the chain of
Sm isotopes
between A=128 and A=188.}%
\label{fig4}%
\end{figure}

\begin{figure}[ptb]
\includegraphics[width=3.1in]{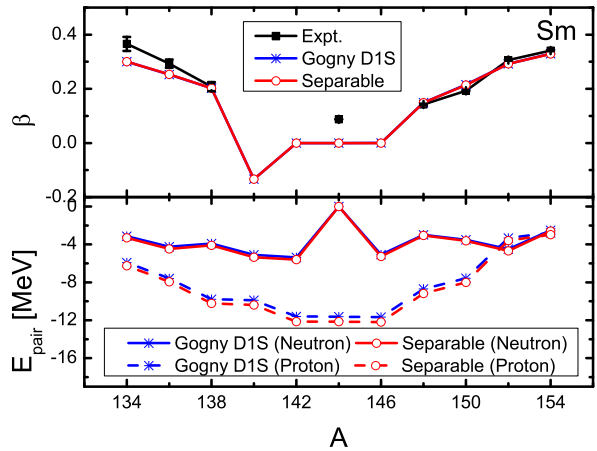}\caption{(Color online) Comparison
between RHB calculations based on NL3 with the original Gogny force
D1S (stars) and with the new separable force adjusted to Gogny D1S
(empty circles). Upper panel: Dependence of the deformation $\beta$
for a chain of Sm isotopes between A=134 and A=154. Lower panel: the
corresponding pairing
energies}%
\label{fig5}%
\end{figure}

We also perform RHBZ calculations with the separable pairing
interaction for a chain of Sm (Z=62) isotopes in the rare-earth
region. Several shape transitions are expected along this isotope
chain. The calculated ground state properties of Sm-isotopes,
especially the deformations are shown in Figs.~\ref{fig3} ,
\ref{fig4}, and \ref{fig5}

In the top panel of Fig.~\ref{fig3} we plot the total binding energy
for Sm-isotopes as a function of neutron number for $66 \leq N
\leq102$. In comparison, we also show results obtained in two other
theoretical models: (i) the RMF model with the parameter set
NL3~\cite{LRR.99} in the ph-channel and the pairing correlations
included by the Bardeen-Cooper-Schrieffer (BCS) formalism with
constant pairing gaps obtained from the prescription of
Ref.~\cite{MN.92}, and (ii) non-relativistic HFB
calculations~\cite{HG.07} with the Gogny force D1S~\cite{BGG.91}. In
the lower panel of Fig.~\ref{fig3} we display the discrepancy of the
binding energies obtained in these three models from the
experimental data~\cite{AWT.03}. All the calculated binding energies
are in good agreement with the experimental data. We observe that
our results obtained from RHBZ-calculations with the new separable
pairing interaction deviate from the experimental binding energies
by less than 0.2\% and that they are for many cases in slightly
better agreement with experiment than the other models

The deformations and shapes of nuclei play a crucial role in
defining the properties such as nuclear sizes and isotopes shifts.
They are strongly affected by the pairing correlations. In the
Fig.~\ref{fig4} we show the quadrupole deformation parameter $\beta$
derived from Eq.~(\ref{E13}) for the Sm isotopes between $N$=66 and
$N$=126. We find that the deformations obtained from RHBZ
calculations with the new separable pairing force are very close to
the results from non-relativistic HFB-calculations with the Gogny
force D1S and in good agreement with the experimental
data~\cite{RNT.01}. Although simple RMF + BCS theory provides a
reasonable description for the binding energies, the deformation
parameters of neutron-rich nuclei calculated in this model deviate
slightly from the experimental data in comparison with those
obtained for RHBZ and HFB theory.

Finally we show in Fig. \ref{fig5} that there is excellent agreement
between the calculations with the separable force presented in this
paper and the original Gogny force in the pairing channel. The
resulting deformation parameters shown in the upper panel are
identical and for the pairing energies in the lower panel the
differences are negligible.

\section{summary}

We have presented first results obtained by axially symmetric
relativistic Hartree-Bogoliubov calculations using the new separable
pairing force introduced in Ref.~\cite{TMR.09a}. This separable
force is translational invariant and has finite range. It contains
two parameters which are adjusted to reproduce the bell shape curve
of the pairing gap at the Fermi surface obtained from the Gogny
force in nuclear matter. In these RHBZ calculations for finite
nuclei the two-body matrix elements of this force are not exactly
separable because of translational invariance. However, using one
and two dimensional Talmi-Moshinsky brackets, they can be evaluated
in an anisotropic axially symmetric harmonic oscillator basis as a
sum of separable terms. We investigate the convergence properties of
this series for various deformations, in particular for normal
prolate, for normal oblate and for super deformed cases. We find
that the expansion of the pairing matrix elements converges
relatively well and that an appropriate truncation provides an
excellent approximation. This allows a considerable reduction of
computing recourses such as time and memory in practical
applications of the RHBZ theory. In particular we study the ground
state properties of a chain of well deformed Sm-isotopes within this
model. We find excellent agreement of our results with those
obtained by using the non-relativistic HFB theory with the Gogny
force D1S, the RMF + BCS theory based on the constant pairing gap
approximation, and with the available experimental data.

Results obtained by RHB-theory with the original Gogny force D1S and
with the separable force derived from it are basically identical.
Therefore we can conclude that this simple pairing interaction can
be applied in future applications of the RHBZ approach in nuclei far
from stability instead of the complicated Gogny force.

\begin{acknowledgments}
We are grateful to Luis Robledo for valuable discussions on the
properties of oscillator functions in various dimensions. This
research was supported by the National Natural Science Foundation of
China under Grants 10875150, 10775183, and 10535010; the Major State
Basis Research Development of China under Contract 2007CB815000; by
the Bundesministerium f\"{u}r Bildung und Forschung (BMBF), Germany,
under Project 06 MT 246, and by the DFG cluster of excellence
\textquotedblleft Origin and Structure of the
Universe\textquotedblright\ (www.universe-cluster.de).
\end{acknowledgments}


\end{document}